\newif\ifpreprint
\preprinttrue

\ifpreprint
\documentclass[nofootinbib,aps,prd,preprint,12pt]{revtex4}
\else
\documentclass[aps,prl,twocolumn]{revtex4}
\fi

\usepackage{amsmath,amsfonts}

\def\bra#1#2{[#1\,#2]}
\def\ket#1#2{\langle #1\,#2\rangle}

\begin{document}

\ifpreprint
\hfill Brown-HET-1587
\vskip 1 cm 
\fi

\title{From Twistor String Theory To Recursion Relations }

\author{Marcus Spradlin and Anastasia Volovich}

\affiliation{Brown University, Providence, Rhode Island 02912, USA}

\begin{abstract}
Witten's twistor string theory gives rise to an enigmatic formula
\ifpreprint
[hep-th/0403190]
\fi
known as
the ``connected prescription'' for tree-level Yang-Mills
scattering amplitudes.
We derive a link representation for the
connected prescription by Fourier transforming it to mixed
coordinates in terms of both twistor and dual twistor variables.
We show that
it can be related to other representations of amplitudes by applying
the global residue theorem to deform the contour of integration.
For
six and seven particles we demonstrate explicitly that certain
contour deformations rewrite the connected prescription as the BCFW
representation, thereby establishing a concrete link
between Witten's twistor string theory and the dual formulation for
the $S$-matrix of ${\mathcal{N}} = 4$ SYM recently
proposed by Arkani-Hamed et.~al.
Other choices of integration contour also give rise to
``intermediate prescriptions''.
We expect a similar
though more intricate structure for more general amplitudes.
\end{abstract}

\maketitle

\section{Introduction}

Witten's twistor string theory proposal~\cite{Witten:2003nn}
launched a series of developments
which have greatly expanded our understanding of the mathematical
structure of scattering amplitudes over the past
several years, particularly in maximally supersymmetric
Yang-Mills theory (SYM).
The most computationally useful technology to have emerged
from subsequent developments is the
Britto-Cachazo-Feng-Witten (BCFW) on-shell
recursion relation~\cite{Britto:2004ap,Britto:2005fq}, the discovery
of which initiated a vast new industry for the computation
of amplitudes.
Building
on~\cite{Hodges},
two recent papers~\cite{Mason:2009sa,ArkaniHamed:2009si}
have paved the way for a return to twistor
space by showing that the BCFW recursion has a natural
formulation there.
Here we bring this set of developments full circle by demonstrating
a beautiful connection between
the original twistor string proposal and
the dual formulation for the $S$-matrix of SYM recently proposed
by Arkani-Hamed et.~al.~\cite{ArkaniHamed:2009dn}.
In particular we show a concrete 
relation between the former and the BCFW representation of
amplitudes.

Our specific focus is on the connected
prescription~\cite{Roiban:2004yf} due to Roiban and the authors
(see also~\cite{Roiban:2004vt,Roiban:2004ka,
Spradlin:2005hi,Vergu:2006np}),
a fascinating but mysterious formula which has been
conjectured to encode
the entire tree-level $S$-matrix of SYM:
\begin{equation}
\label{eq:Tformula}
{\cal T}_{n,k}({\cal Z}) = \int [d {\cal P}]_{k-1}
d^n \sigma \prod_{i=1}^n \frac{\delta^{3|4}({\cal Z}_i - {\cal P}(\sigma_i))}
{\sigma_i - \sigma_{i+1}}.
\end{equation}
Here ${\cal P}(\sigma)$ denotes a ${\mathbb{P}}^{3|4}$-valued polynomial
of degree $k-1$ in $\sigma$ and $[d{\cal P}]_{k-1}$ is the natural
measure on the space of such polynomials.
We review further details shortly but pause to note that
this formula simply expresses the content of
Witten's twistor string theory: the N${}^{k-2}$MHV superamplitude
is computed as the integral of an
open string current algebra correlator over the moduli space of
degree $k-1$ curves in supertwistor space ${\mathbb{P}}^{3|4}$.

The formula~(\ref{eq:Tformula})
manifests several
properties which scattering amplitudes must possess, including
conformal invariance and cyclic
symmetry of the superamplitude, both of which are hidden
in other representations such as BCFW.  It is also relatively easy
to show that it possesses the correct soft and collinear-particle
singularities, as well as (surprisingly) parity
invariance~\cite{Roiban:2004yf,Witten:2004cp}.
Despite these conceptual strengths the connected prescription has
received relatively little attention
over the past five years because
it has resisted attempts to relate it directly to the more
computationally useful BCFW recursion relation.

Here we remedy this situation by showing for the first time
a direct and beautiful relation between the connected
prescription~(\ref{eq:Tformula}) and the BCFW recursion.
Specifically we demonstrate explicitly
for $n=6,7$
(and expect a similar though more intricate story for general $n$)
that different choices of integration contour
in~(\ref{eq:Tformula})
compute different, but equivalent, representations of tree-level
amplitudes~\footnote{It has been argued
in~\cite{Gukov:2004ei} that the connected prescription
can also be related to the CSW representation~\cite{Cachazo:2004kj}
by a contour deformation in the moduli space of curves.
}.  The privileged contour singled out by the
delta-functions appearing in~(\ref{eq:Tformula}) computes the
connected prescription representation in which the $n$-particle
N${}^{k-2}$MHV amplitude is expressed as a sum of residues of the integrand
over
the roots of a
polynomial of degree $\genfrac{<}{>}{0pt}{}{n-3}{k-2}$
(where $\genfrac{<}{>}{0pt}{}{a}{b}$ are Eulerian numbers).
Different representations of tree-level amplitudes, including
BCFW representations as well as intermediate prescriptions
similar to those of~\cite{Gukov:2004ei,Bena:2004ry}, are all
apparently encoded in various residues of the integrand ${\cal T}_{n,k}$ and
are computed by choosing various appropriate contours.
The equivalence of different representations follows from the global
residue theorem, a multidimensional analogue of Cauchy's theorem.

The integrand ${\cal T}$
has many residues in common with
\begin{equation}
\label{eq:Lformula}
{\cal L}_{n,k}({\cal W}) = \int
[dC]_{k \times n}
\prod_{i=1}^n
\frac{\delta^{4|4}(C_{\alpha i} {\cal W}_i)}{(i,i+1,\ldots,i+k-1)}
\end{equation}
recently written down by
Arkani-Hamed et.~al.~\cite{ArkaniHamed:2009dn}.
Here $[dC]_{k \times n}$ is the measure on the space of $k \times n$
matrices modulo left-multiplication by $GL(k)$ and
$(m_1,\cdots,m_k)$ denotes
the minor obtained from $C$ by keeping only columns $m_1,\ldots,m_k$.
Residues of both ${\cal T}$ and ${\cal L}$ compute BCFW representations of
tree amplitudes.  In addition,
${\cal T}$ also computes various other tree-level
representations while ${\cal L}$ evidently computes parity-conjugate
P(BCFW) representations at tree-level as well as leading singularities
of loop amplitudes.
It is natural to wonder whether there exists some
richer object
${\cal D}$ (for ``dual'') which contains information
about various connected and disconnected representations of amplitudes
at tree level and at all loops.
This could help shed further light on twistor string theory at the loop level.

It is not yet known which contour computes which object from the integrand
${\cal L}$.
In contrast, as mentioned above, the connected prescription ${\cal T}$
comes equipped with
a certain privileged contour which calculates the tree
amplitude.  Various other contours which compute different
representations of the same amplitude
can be easily determined by applying
the global residue theorem.
We hope that a better understanding of the relation between ${\cal L}$
and ${\cal T}$ may allow us to transcribe information about the privileged
contour from the latter to the former.

\ifpreprint
In section 2 we review the connected prescription for computing
scattering amplitudes and derive its link representation by Fourier
transforming it to mixed ${\cal Z}$, ${\cal W}$ variables.
In section 3 we demonstrate the precise relation between the connected
prescription, BCFW and intermediate representations of all
six- and seven-particle
amplitudes.
\fi

\section{Linking The Connected Prescription}

Let us begin by reviewing some details of the connected
prescription formula~(\ref{eq:Tformula}) for the color-stripped $n$-particle
N${}^{k-2}$MHV scattering amplitude.
The $4|4$ component homogeneous coordinates for the $i$-th
particle in ${\mathbb{P}}^{3|4}$ are
${\cal Z}_i = (\lambda_i^\alpha, \mu_i^{\dot{\alpha}}, \eta_i^A)$ with
$\alpha,\dot{\alpha} = 1,2$ and $A =1,2,3,4$.
In split signature $--++$ the spinor helicity variables
$\lambda_i^\alpha, \widetilde{\lambda}_i^{\dot{\alpha}}$ can be
taken as independent real variables and the twistor transform realized
in the naive way as a Fourier transform from
$\widetilde{\lambda}_i^{\dot{\alpha}}$ to $\mu_i^{\dot{\alpha}}$.

As emphasized in~\cite{Roiban:2004yf}
(see also~\cite{Vergu:2006np})
the integral~(\ref{eq:Tformula}) must be
interpreted as a contour integral in a multidimensional
complex space.  The delta functions specify
the contour of integration according to the usual rule
\begin{equation}
\int d^m z \ h(\vec{z}) \prod_{i=1}^m \delta(f_i(\vec{z}))
= \sum
h(\vec{z})
\left[ \det \frac{\partial f_i}{\partial z_j} \right]^{-1}
\end{equation}
with the sum taken over the set of $\vec{z}_*$ satisfying
$f_1(\vec{z}_*) = \cdots = f_m(\vec{z}_*) = 0$.
In practice the calculation of any $n$-particle
N${}^{k-2}$MHV amplitude therefore reduces to the problem
of solving certain polynomial equations which appear to have
$\genfrac{<}{>}{0pt}{}{n-3}{k-2}$ roots in general.

To write the connected formula slightly more explicitly we
first express the delta functions on ${\mathbb{P}}^{3|4}$
in terms of homogeneous coordinates via the contour
integral
\begin{equation}
\delta^{3|4}({\cal Z} - {\cal Z}') = 
\int \frac{d\xi}{\xi} \delta^{4|4}({\cal Z} - \xi
{\cal Z}').
\end{equation}
Next we parameterize the degree $k-1$ polynomial
${\cal P}$ in terms of its $k$ ${\mathbb{C}}^{4|4}$-valued
supercoefficients
${\cal A}_d$ as
\begin{equation}
{\cal P}(\sigma) = \sum_{d=0}^{k-1} {\cal A}_d \sigma^d.
\end{equation}
Using these ingredients~(\ref{eq:Tformula}) may be expressed as
\begin{equation}
\label{eq:integralone}
{\cal A}({\cal Z}) = \int \frac{d^{4k|4k} {\cal A}\, d^n \sigma\,
d^n \xi}{{\rm vol}\,GL(2)}
\prod_{i=1}^n \frac{ \delta^{4|4}({\cal Z}_i -
\xi_i {\cal P}(\sigma_i))}{\xi_i(\sigma_i-\sigma_{i+1})},
\end{equation}
where we have indicated that the integrand and measure
are invariant under a GL(2) acting as M\"obius
transformations of the $\sigma_i$ combined with a simultaneous
compensating reparameterization of the curve ${\cal P}(\sigma)$.
This symmetry must be gauge-fixed in the usual way.

Motivated by~\cite{ArkaniHamed:2009si}
we now consider expressing the connected
prescription~(\ref{eq:Tformula}) in a mixed representation where
some of the particles are specified in terms of the ${\cal Z}$ variables
as above while others are specified in terms of
the variables ${\cal W} = (\widetilde{\mu}^{\dot{a}},
\widetilde{\lambda}^{\dot{a}}, \widetilde{\eta}_{{A}})$ related by Fourier transform
\begin{equation}
{\cal F}({\cal W}) = \int d^{4|4} {\cal Z} \ F({\cal Z})
\,e^{i {\cal W} \cdot {\cal Z}},
\end{equation}
where
${\cal W} \cdot {\cal Z} =\widetilde{\mu}\cdot \lambda - \mu \cdot\widetilde{\lambda} + \eta \cdot \widetilde{\eta}$.
A particularly convenient choice for the N${}^{k-2}$MHV amplitude
is to leave precisely $k$ particles in terms of ${\cal Z}$
and transform the rest
to ${\cal W}$.  This replaces the $4n|4n$ delta-functions
in~(\ref{eq:integralone}) with
\begin{equation}
\prod_i \exp \left( i \xi_i {\cal W}_i \cdot {\cal P}(\sigma_i) \right)
\prod_J \delta^{4|4}({\cal Z}_J - \xi_J {\cal P}(\sigma_J)).
\end{equation}
Here and in all that follows it is implicit that sums or products over $i$
run over the subset of the $n$ particles expressed in the ${\cal W}$
variables while sums or products over $J$ run over the particles
expressed in terms of ${\cal Z}$'s.

The utility of our choice is that there are now precisely as many
delta functions as supermoduli ${\cal A}$, which moreover can be
integrated out trivially since they appear
linearly inside delta functions.
This operation sets
\begin{equation}
{\cal P}(\sigma) = \sum_J \frac{{\cal Z}_J}{\xi_J} \prod_{K \ne J}
\frac{\sigma_K - \sigma}{\sigma_K - \sigma_J}
\end{equation}
which is easily seen to satisfy ${\cal P}(\sigma_J) = {\cal Z}_J/\xi_J$.
The resulting expression for the integral may be cleaned up
with the help of the change of variables
\begin{equation}
x_i = \xi_i \prod_K (\sigma_K - \sigma_i), \qquad
x_J^{-1} = \xi_J \prod_{K \ne J} (\sigma_K - \sigma_J).
\end{equation}
which (ignoring for the moment overall signs)
transforms~(\ref{eq:integralone})
into
an integral which can be put
into the form of a link representation
\begin{equation}
\label{eq:link}
{\cal A}({\cal W}_i, {\cal Z}_J) = \int dc_{iJ}\ U(c_{iJ})
\,e^{i c_{iJ} {\cal W}_i \cdot {\cal Z}_J}
\end{equation}
(as introduced in~\cite{ArkaniHamed:2009si}) with the integrand
given by
\begin{equation}
\label{eq:Udef}
U(c_{iJ}) = \int \prod_{a=1}^n
\frac{d\sigma_a\,dx_a}{x_a(\sigma_a - \sigma_{a+1})} \prod_{i,J}
\delta\left( c_{iJ} - \frac{x_i x_J}{\sigma_J - \sigma_i}\right).
\end{equation}

Note that this expression still requires $GL(2)$
gauge fixing.
Usually this is accomplished by freezing four variables
$\sigma_1,\sigma_2,\sigma_3,x_1$
to arbitrary values with the Jacobian
\begin{equation}
\int d\sigma_1\,d\sigma_2\,d\sigma_3\,dx_1 =
x_1 (\sigma_1-\sigma_2)(\sigma_2-\sigma_3)(\sigma_3-\sigma_1).
\end{equation}
Consequently in~(\ref{eq:Udef}) there are effectively only $2n-4$ integration
variables and $k (n-k-4)$ delta functions, so that after integrating
out the $x$'s and $z$'s there remain in $U$ a net $(k-2) (n-k-2)$
delta functions.

As emphasized in~\cite{ArkaniHamed:2009si}
an important feature of the link representation is that
returning physical space is simple because the Fourier transforms
$\mu_i^{\dot{\alpha}} \to \widetilde{\lambda}_i^{\dot{\alpha}}$,
$\widetilde{\mu}_i^\alpha \to \lambda_i^\alpha$ turn the exponential
factors
in~(\ref{eq:link}) into
\begin{equation}
\label{eq:physdelta}
\prod_i \delta^2 (\lambda_i^\alpha - c_{iJ} \lambda_J^\alpha)
\prod_J \delta^2 (\widetilde{\lambda}_J^{\dot{\alpha}} + c_{iJ}
\widetilde{\lambda}_i^{\dot{\alpha}}).
\end{equation}
For given kinematics $(\lambda_i^\alpha,
\widetilde{\lambda}_i^{\dot{\alpha}})$ these equations
fix the $k(n-k)$ $c_{iJ}$ as linear functions of $(k-2)(n-k-2)$
remaining free parameters denoted $\tau_\gamma$.
Finally we obtain the physical space
amplitude in terms of $U$ as
\begin{equation}
\label{eq:pstransform}
{\cal A}(\lambda,\widetilde{\lambda})
= 
J\, \delta^4( {\textstyle{\sum}} p_i )
\int d^{(k-2)(n-k-2)} \tau\ U(c_{iJ}(\tau_\gamma)),
\end{equation}
where $J$ is the Jacobian from integrating out~(\ref{eq:physdelta}).
We will always implicitly choose for simplicity a parameterization
of $c_{iJ}(\tau_\gamma)$ for which $J=1$.
Before proceeding let us again emphasize
that each $c_{iJ}(\tau_\gamma)$ is linear in the $\tau$'s.

\section{Examples}

For the trivial case of MHV amplitudes ($k=2$) the remaining integrations
are easily carried out, leading to
\begin{equation}
U^{--+\cdots+} =
\frac{1}{c_{31} c_{n2}} \prod_{i=3}^{n-1}
\frac{1}{c_{i,i+1:1,2}},
\end{equation}
in terms of
$c_{ij;KL} = c_{iK} c_{jL} - c_{iL} c_{jK}$.
The $\overline{\rm MHV}$ case $k=n-2$ yields the same result with
$c_{ab} \to c_{ba}$.  When transformed to physical space
using~(\ref{eq:pstransform})
these yield respectively the Parke-Taylor formula and its conjugate.

\subsection{6-Point Amplitudes}

Next we consider
the six-particle alternating helicity amplitude,
for which we find from~(\ref{eq:Udef})
the representation
\begin{equation}
\label{eq:uuu}
U^{+-+-+-} = \frac{1}{c_{14} c_{36} c_{52}} \delta(S_{135:246})
\end{equation}
where $S$ refers to the sextic polynomial
\ifpreprint
\begin{equation}
S_{ijk:lmn} =
c_{im} c_{jm} c_{kl} c_{kn} c_{ij:ln}
- c_{in} c_{jn} c_{kl} c_{km} c_{ij:lm}
- c_{il} c_{jl} c_{km} c_{kn} c_{ij:mn}.
\end{equation}
\else
\begin{multline}
S_{ijk:lmn} =
c_{im} c_{jm} c_{kl} c_{kn} c_{ij:ln}
\\
- c_{in} c_{jn} c_{kl} c_{km} c_{ij:lm}
- c_{il} c_{jl} c_{km} c_{kn} c_{ij:mn}.
\end{multline}
\fi
In this example the
appearance of $\delta(S_{135:246})$ can be understood as follows:
we are trying to express nine variables $c_{iJ}$ in terms of
eight variables (the $x$'s and $z$'s) by solving the delta-function
equations
\begin{equation}
c_{iJ} = \frac{x_i x_J}{\sigma_J - \sigma_i}.
\end{equation}
A solution to this overconstrained set of equations for the $c_{iJ}$
exists if and only
if the sextic $S_{135:246}$ vanishes.

{}From~(\ref{eq:uuu}) we arrive at the expression
\begin{equation}
\label{eq:six}
A^{+-+-+-} = \int d\tau\ 
\frac{1}{c_{14} c_{36} c_{52}} \delta(S_{135:246})
\end{equation}
for the physical space amplitude.
In this case $S_{135:246}$ is quartic in the single $\tau$ parameter.
By choosing numerical values for the external kinematics and summing
over the four roots of $S_{135:246}$ one can verify that~(\ref{eq:six})
reproduces the correct amplitude.

Now consider more generally the object
\begin{equation}
\label{eq:integrand}
\frac{1}{c_{14} c_{36} c_{52}} \frac{1}{S_{135:246}}
\end{equation}
as a function of $\tau$.  The contour integral of this object around
the four zeroes of $S_{135:246}$ evidently computes the alternating
helicity six-particle amplitude.  But~(\ref{eq:integrand})
has three other poles located at the vanishing of
$c_{14}$, $c_{36}$
or $c_{52}$.  By Cauchy's theorem
we know that the sum of these three residues
computes minus the amplitude,
\begin{equation}
A^{+-+-+-} = - \int d\tau \frac{1}{S_{135:246}} \delta(c_{14} c_{36} c_{52}).
\end{equation}
Since the $c_{iJ}$ are linear in $\tau$ it is simple to calculate the
corresponding residues analytically, and one obtains
\ifpreprint
\begin{equation}
\frac{\bra{1}{3}^4 \ket{4}{6}^4}{\bra{1}{2} \bra{2}{3} \ket{4}{5} \ket{5}{6} s_{123}
\langle 6 | 5 + 4|3 ] \langle 4|5 + 6| 1]}
+ (i \to i + 2) + (i \to i + 4)
\end{equation}
\else
\begin{multline}
\frac{\bra{1}{3}^4 \ket{4}{6}^4}{\bra{1}{2} \bra{2}{3} \ket{4}{5} \ket{5}{6} s_{123}
\langle 6 | 5 + 4|3 ] \langle 4|5 + 6| 1]}
\\
+ (i \to i + 2) + (i \to i + 4)
\end{multline}
\fi
which is the BCFW representation for the amplitude\!

Analysis of the other two independent six-particle helicity configurations 
proceeds along the same lines with link representations obtained
from~(\ref{eq:Udef}):
\begin{eqnarray}
U^{+++---} &=& \frac{c_{25}}{
c_{12:45} c_{23:56}
} \delta(S_{123:456}),
\\
U^{++-+--} &=& \frac{c_{16}}{
c_{13} c_{46} c_{12:56}
} \delta(S_{124:356}).
\end{eqnarray}
In each case the connected presentation expresses the amplitude as
a sum over the four roots of the quartic
$S_{ijk:lmn}$ in the $\tau$-plane, which
a simple application of Cauchy's theorem relates to
a sum over simple
linear roots which compute the BCFW representation of the amplitude.

\subsection{7-Point Amplitudes}

For the seven-particle split helicity amplitude we find
\begin{equation}
\label{eq:sevenlink}
U^{++++---} =
\frac{c_{25} c_{26} c_{36} c_{37}}{c_{12:56} c_{34:67}}
\delta(S_{123:567}) \delta(S_{234:567}).
\end{equation}
There are now two $\tau$ variables, and the locus
where both of the delta functions vanish consists of 14 isolated
points in ${\mathbb{C}}^2$.  The coordinates of these
points are determined by the vanishing of a polynomial which
is a product of one of degree 11 and three of degree 1.
The three linear roots do not contribute because the numerator
factors in~(\ref{eq:sevenlink}) vanish there.
Therefore~(\ref{eq:sevenlink}) represents the amplitude as a sum over the roots of
a degree 11 polynomial, as expected for the connected prescription
for $n=7$, $k=3$.

To proceed
we must use the multidimensional
analog of Cauchy's theorem known as the global residue theorem:
\begin{equation}
\label{eq:grt}
\oint_{f_1 = \cdots = f_n = 0} d^nz\ \frac{h(z)}{f_1(z) \cdots f_n(z)} = 0
\end{equation}
when $h(z)$ is a polynomial of degree
less than $\sum \deg f_i - (n + 1)$, so that it has no poles at finite $z$
and the integrand falls off sufficiently fast to
avoid a pole at infinity.

To apply~(\ref{eq:grt}) to~(\ref{eq:sevenlink})
we consider the integrand
\begin{equation}
\frac{c_{25} c_{26} c_{36} c_{37}}{c_{12:56} c_{34:67}}
\frac{1}{S_{123:567} S_{234:567}}.
\end{equation}
There are seven independent ways of grouping the terms in the denominator
into a product $f_1 f_2$.
The choice
\begin{equation}
f_1 = c_{12:56} S_{234:567}, \qquad
f_2 = c_{34:67} S_{123:567}
\end{equation}
is particularly nice:  in this application of
the global residue theorem all 11 poles at the locus
$S_{123:567} = S_{234:567} = 0$ contribute as do the roots located at
\begin{eqnarray}
c_{12:56} = S_{123:567} &=& 0, \\
c_{34:67} = S_{234:567} &=& 0, \\
c_{12:56} = c_{34:67} &=& 0,
\end{eqnarray}
which amazingly turn out to each consist of a single linear root.
The global residue theorem expresses the connected representation of the
amplitude as (minus) the sum of these three linear roots, which a simple
calculation reveals as precisely the three terms contributing to the
BCFW representation of the amplitude.

Equally amazing is the choice
\begin{equation}
f_1 = S_{123:567}, \qquad
f_2 = c_{12:56} c_{34:67} S_{234:567}.
\end{equation}
This contour computes the sum of residues at 15 poles; 11 of those
are the connected prescription poles which we know compute the correct
physical amplitude, while the others consist of a single linear root
together with four quartic roots.  Schematically then this global
residue theorem identity expresses
\begin{equation}
A^{++++---} = \sum {\rm 11~roots} = - \sum {\rm 4~roots} - {\rm 1~root}.
\end{equation}
We interpret the right-hand side of this equation as an `intermediate'
prescription~\cite{Gukov:2004ei,Bena:2004ry}, obtained by BCFW decomposing
$A^{++++---}$ once into the product of a 3-particle
amplitude with a split-helicity six-particle amplitude, and then computing
the latter via the connected prescription as a sum over four roots.

We end by tabulating link representations for the remaining
independent seven-particle helicity amplitudes
\begin{equation}
\begin{split}
U^{+++-+--} &=
\frac{c_{26} c_{27} c_{25:46}}{c_{12:46} c_{23:67}}
\delta(S_{125:467}) \delta(S_{235:467}),
\\
U^{++-++--} &=
\frac{c_{23} c_{56} c_{57} c_{25:36}}{c_{53} c_{12:36} c_{45:67}}
\delta(S_{125:367}) \delta(S_{245:367}),
\\
U^{++-+-+-} &=
\frac{c_{17} c_{43} c_{14:57}}{c_{47} c_{63} c_{12:57}}
\delta(S_{124:357}) \delta(S_{146:357}).
\end{split}
\end{equation}
As usual we interpret $\delta(u) = 1/u$ in the integrand
with the delta functions indicating the preferred contour
which computes the connected prescription representation of the
amplitude.

\section*{Acknowledgments}

We are grateful to N.~Arkani-Hamed and F.~Cachazo
for extensive discussions and
enormous encouragement and to C.~Vergu and C.~Wen for helpful comments.
This work was supported in part by the
Department of Energy under contract DE-FG02-91ER40688 Task J OJI (MS)
and Task A (AV),  the National Science Foundation under grants
PHY-0638520 (MS), PECASE PHY-0643150 (AV) and ADVANCE 0548311 (AV).

\end{document}